\documentclass[11pt,a4paper]{article}

\usepackage[utf8]{inputenc}
\usepackage[T1]{fontenc}
\usepackage{lmodern}
\usepackage[english]{babel}
\usepackage{geometry}
\usepackage{amsmath,amssymb,amsthm}
\usepackage{graphicx}
\usepackage{booktabs}
\usepackage{microtype}
\usepackage{xcolor}
\usepackage[round,authoryear]{natbib}
\usepackage{hyperref}
\hypersetup{
  colorlinks=true,
  linkcolor=black,
  citecolor=black,
  urlcolor=blue
}
\usepackage{caption}
\usepackage{authblk}
\newtheorem{hypothesis}{Hypothesis}

\title{Social Policy of Large Language Models:\\ How GPT, Claude, DeepSeek and Grok Allocate Social Budgets in Spain and Germany}

\author[1]{Claudia Benavides Cantos}
\author[1,2]{Eduardo C. Garrido-Merch\'an\thanks{Corresponding author: \texttt{ecgarrido@icade.comillas.edu}}}
\affil[1]{Universidad Pontificia Comillas, Faculty of Economics and Business Administration, Madrid, Spain}
\affil[2]{Institute for Research in Technology (IIT), Universidad Pontificia Comillas, Madrid, Spain}

\date{}

\begin{document}

\maketitle

\begin{abstract}
\noindent We study how four widely used large language models, namely Claude, GPT-4o, DeepSeek and Grok, distribute a fixed national social budget across twelve macro-areas of public expenditure under two European national contexts, Spain and Germany. Each combination of model and country is queried six times under identical prompts and generation parameters, producing forty-eight independent allocations that are compared against approximate Organisation for Economic Co-operation and Development (OECD) reference budgets and against each other. We formalise five hypotheses regarding geopolitical bias, housing under-allocation, structural convergence, sensitivity to national context, and under-representation of politically sensitive categories. The differences between models are then validated through Kruskal--Wallis tests on each macro-area, with post-hoc Mann--Whitney U comparisons under Bonferroni correction, and complemented by an analysis of pairwise Pearson correlations and a lexical examination of the textual justifications produced by each model. The results show that all four models share a systematic implicit social policy that diverges from real European spending structures: pensions are under-allocated by a factor close to three, while housing and employment are over-allocated by factors of four and two respectively. The principal axis of differentiation between models is not geopolitical, since Claude and DeepSeek are the most correlated pair across both countries, but rather a contrast between concentration and dispersion of the budget. Only Claude exhibits substantive sensitivity to the national context. The conclusions delimit the conditions under which language models may responsibly support, but not replace, expert deliberation in public budgeting.\\

\end{abstract}

\section{Introduction}

Large language models have become one of the most disruptive technological developments in artificial intelligence in the last decade. Built on the transformer architecture introduced by \citet{vaswani2017attention}, they are capable of generating coherent natural language, performing complex reasoning tasks for which they were not explicitly trained, and adapting to new instructions through in-context learning. Their ease of use and flexibility have driven their rapid adoption across business services, education, healthcare, and, increasingly, public administration and policy-related decision support \citep{ortiz2024economia}.

The expansion of these systems into the public sphere raises the central question of this work. Governments face growing social challenges, ranging from unemployment and poverty to housing shortages and insufficient access to essential services, against budgets that are inherently constrained. Within this context, large language models have begun to be explored as potential support tools for public decision making, for example in policy simulation, analysis of legislative texts, and the elaboration of alternative budget proposals. Yet their training corpora encode the biases of the people, institutions and discursive traditions that produced them. As a consequence, ideological biases may seep into the models and reinforce inequality through their responses \citep{okerlund2022chatterbox}. More worrying still, recent research has shown that users who interact with a biased model tend to shift their own opinions, and even their decisions about the distribution of public resources, in the direction of the model's bias, independently of their own political affiliation and largely without being aware of it \citep{fisher2024biased}.

Although academic interest in bias in language models is rising, systematic and quantitative analyses of how different models behave when confronted with concrete problems of allocating public resources are scarce \citep{echterhoff2024cognitive}. What is at stake is not only the final allocation, but also the reasoning behind each response, the priorities that the model implicitly reveals, the differences between models, and the coherence of the proposed allocation with the real budgetary structure of each country. This empirical gap motivates the present study at the intersection of artificial intelligence and social policy.

We therefore examine the way in which four widely used commercial language models, Claude developed by Anthropic, GPT-4o developed by OpenAI, DeepSeek developed by DeepSeek AI, and Grok developed by xAI, distribute a hypothetical social budget across twelve macro-areas of public expenditure for two European national contexts, Spain and Germany. The two countries belong to the European Union, where social protection expenditure averaged 27.3\% of gross domestic product in 2024 \citep{eurostat2025benefits}, yet they exhibit distinct welfare structures that make them complementary cases for a comparative analysis. The four models were selected to combine systems developed in different geographical and cultural environments, both United States-based and Chinese, thereby introducing a geopolitical dimension that has become salient in the contemporary debate over artificial intelligence \citep{rozado2024political}.

The contribution of this work is threefold. First, we propose a controlled experimental protocol that fixes the prompt, the model parameters and the taxonomy of expenditure across all calls, so that observed differences can be attributed to the model as the principal independent variable. Second, we report a statistical analysis based on the Kruskal--Wallis test on each macro-area, with post-hoc pairwise Mann--Whitney U tests under Bonferroni correction, complemented by mean absolute deviation against OECD reference structures and by pairwise Pearson correlations between models. Third, we discuss a lexical analysis of the textual justifications generated by each model, which reveals two complementary rhetorical regimes, one organised around monetary quantification and another around programmatic content, and connects the quantitative findings with the implicit normative frames invoked by the models.

The remainder of the article is organised as follows. Section~\ref{sec:related} situates the contribution within the literature on the political preferences of language models, their use as classifiers and assistants in social processes, their role as agents in simulations of human behaviour, and their deployment in deliberative and democratic settings. Section~\ref{sec:methods} describes the experimental design, the taxonomy of expenditure, the prompt and the statistical apparatus. Section~\ref{sec:results} reports the empirical findings. Section~\ref{sec:discussion} discusses their implications for the use of language models as decision support in public budgeting. Section~\ref{sec:conclusion} concludes and outlines directions for future work.

\section{Related work}\label{sec:related}

Research on large language models has expanded rapidly into the political and social domain, yet decisions about public budget allocation have only been examined indirectly. We organise the relevant literature into four complementary strands.

The first strand concerns the political preferences of language models themselves. The most systematic analysis is provided by \citet{rozado2024political}, who applied eleven established political orientation tests to twenty-four conversational systems, including closed models such as GPT-3.5, GPT-4 and Gemini, and open models such as Llama 2 and Mistral. Models that have undergone supervised fine-tuning and reinforcement learning from human feedback exhibited a left-leaning inclination, while unaligned base models did not respond consistently, confirming that the alignment process is the principal determinant of the observed political orientation. \citet{fisher2024biased} complemented this finding with two experiments in which participants interacted with a liberal-leaning, conservative-leaning, or neutral model and were asked to express opinions on political topics and to distribute a budget across four governmental sectors before and after the interaction. Participants converged towards positions and allocations aligned with the bias of the model, even when these differed from their original preferences, and prior knowledge of artificial intelligence mitigated but did not eliminate this effect. Taken together, these results establish that language models carry implicit political preferences and that exposure to them measurably alters human distributive choices.

A second strand uses language models as classifiers and assistants in social processes. \citet{lemens2025positioning} propose querying GPT-4, Claude and LLaMA on the ideological position of political texts and averaging the responses to obtain an estimate, achieving correlations of approximately $0.90$ with expert-based and crowd-based reference scales and outperforming supervised classifiers trained on large annotated corpora. \citet{lee2025ideology} investigate whether language models can generate synthetic survey data on labour policy attitudes in South Korea, finding that the models reproduce the demographic and ideological structure of respondents with reasonable accuracy but tend to overemphasise ideological orientation on polarised issues. Both studies confirm that language models possess analytical capacity in political contexts while exhibiting systematic limitations associated with their training data.

A third strand treats language models as agents in simulations of social and scientific behaviour. \citet{anthis2025simulations} argue that large language model social simulations are already useful for exploratory and pilot studies, identifying diversity, bias, sycophancy, alienness and generalisation as the principal open challenges. \citet{manning2024automated} extend this perspective by using language models as automated social scientists capable of formulating hypotheses, designing experiments and analysing data through structural causal models, with reasonable success in negotiation and judicial scenarios but with magnitudes of effect that are systematically overestimated. \citet{argyle2023silicon} establish that language models can simulate human samples conditioned on demographic information with high fidelity, while \citet{santurkar2023whose} characterise whose opinions language models tend to reflect after fine-tuning. These contributions explore the use of language models as substitutes for human subjects or as automated researchers, rather than as agents that take concrete decisions in a realistic social setting.

A fourth strand examines language models in democratic and deliberative processes. \citet{cox2024voices} show that GPT-4, LLaMA and Claude can identify complex semantic patterns and produce summaries comparable to expert ones in roughly 80\% of cases, while replicating dominant discursive structures and lacking transparency in their inferential process. \citet{gudino2024augmented} propose the use of language models as digital twins of citizens to predict how individuals would vote on proposals they have not directly evaluated, with predictions that surpass party-based heuristics. None of these studies addresses the role of language models as decision-making agents on urgent social policy questions.

The literature thus establishes that language models carry implicit political preferences, influence human decisions, classify political content with high accuracy, simulate social behaviour and assist in democratic processes. The present work addresses a complementary gap by examining the models as agents that allocate a national social budget across a fine-grained taxonomy, comparing their allocations against real budgetary structures, validating differences statistically, and analysing both the quantitative outcomes and the textual reasoning that accompanies them. Our concern is not whether language models reflect public opinion, but whether their implicit distributive preferences align with the structural reality of European welfare states, and whether they can be safely used as decision support in this domain.

\section{Materials and methods}\label{sec:methods}

The experimental design rests on the principle of controlled comparability. All models receive exactly the same prompt, the same generation parameters and the same access conditions through their respective application programming interfaces, so that the model itself can be isolated as the principal independent variable. Three dimensions of analysis are considered simultaneously. The first dimension concerns implicit priorities, that is, the macro-areas and categories that each model systematically prioritises with respect to the others. The second dimension concerns internal consistency, measured by whether a given model produces stable or volatile allocations under repeated queries, on the basis of six independent runs per combination of model and country. The third dimension concerns sensitivity to the national context, that is, whether the allocations vary coherently between the Spanish and the German scenarios in response to the socioeconomic challenges described in the prompt.

The four models employed are GPT-4o by OpenAI (API identifier \texttt{gpt-4o}), Claude Sonnet 4 by Anthropic (\texttt{claude-sonnet-4-20250514}), DeepSeek-V3 by DeepSeek AI (\texttt{deepseek-chat}), and Grok-4 by xAI (\texttt{grok-4}). Three of them were developed in the United States and one in China, providing geopolitical diversity within the set of commercial systems available at the time of the experiment, April 2026. Anthropic's Claude has been developed with explicit emphasis on safety and alignment through Constitutional AI, in which the model is trained to evaluate, criticise and revise its own responses according to an explicit set of principles \citep{bai2022constitutional, anthropic2024claude}. GPT-4o is the multimodal successor of GPT-4, trained with the transformer architecture and aligned by reinforcement learning from human feedback \citep{openai2023gpt4, ouyang2022instruct}. DeepSeek-V3 uses a Mixture-of-Experts architecture that achieves competitive performance with reduced training cost and is the first open-weight model to rival proprietary Western systems on several benchmarks \citep{deepseek2024v3}. Grok-2 and Grok-4 were released by xAI with advanced reasoning capabilities, although the absence of a formal technical report introduces a transparency limitation in comparison with the other three models \citep{xai2024grok2}.

All four models were queried with an identical system prompt, ``You are a public policy expert'', a generation temperature of 0.7 and a maximum response length of 4{,}000 tokens. A homogeneous temperature ensures a comparable level of generative variability across systems, so that the observed differences in allocation cannot be attributed to differences in stochasticity. Each combination of model and country was executed six times through direct API calls, with a two-second interval between consecutive calls to respect the rate limits of each provider. The full design produces $4 \times 2 \times 6 = 48$ independent allocations.

The taxonomy of expenditure was constructed by adapting the functional classification of the OECD Social Expenditure Database \citep{oecd2024socx} and contains twelve macro-areas and seventy-two sub-categories. The twelve macro-areas are Education (7 sub-categories), Health (11), Housing and homelessness (7), Employment and labour market (7), Gender equality and non-discrimination (7), Children and family (6), Pensions and older people (5), Social inclusion and poverty (4), Community and civic life (5), Justice and legal assistance (4), Environment and social sustainability (4), and Migration and integration (5). The standardised prompt provided, for each country, the total budget in absolute terms and as a percentage of gross domestic product, a brief description of the main socioeconomic challenges, and formatting instructions requiring that the percentages sum exactly to one hundred and that each category be accompanied by the corresponding allocation, the absolute amount and a one- or two-sentence justification. For Spain, the prompt specified a budget of \$390 billion, approximately 26\% of gross domestic product, and described youth unemployment of approximately 28\%, the housing crisis, accelerated population ageing, an underfunded long-term care system, regional inequality, and rural depopulation. For Germany, the prompt specified a budget of \$1.20 trillion, approximately 28\% of gross domestic product, and described demographic ageing, the integration of more than two million refugees, the costs of the energy transition, housing shortages in large cities and the persistent east-west economic divide.

Reference budgets for both countries were constructed from the OECD Social Expenditure Database and from Eurostat's statistics on social protection \citep{oecd2024socx, oecd2024society, eurostat2025pension}. Approximately 40\% to 45\% of European social spending is concentrated in pensions and older people, followed by 25\% to 30\% in health, 8\% to 10\% in family and children, 5\% to 6\% in unemployment, 6\% to 8\% in disability, and 1\% to 3\% in housing, with relatively consistent structure across countries. Deviations from these ranges, for instance an allocation of 30\% to housing combined with 2\% to pensions, are interpreted as signals of distributive bias in the proposals generated by the models. Spain devotes approximately 26\% of gross domestic product to social spending, slightly below the European average, with pensions accounting for more than two thirds of social benefit expenditure and roughly 12.6\% of gross domestic product, and with a particularly low share devoted to housing \citep{bde2022gasto, ine2024contabilidad}. Germany devotes approximately 28\% of gross domestic product to social spending, with public pensions as the largest single item, in line with the European average \citep{imf2024weo, eurostat2025pension}.

We formalise five testable hypotheses motivated by the literature.

\begin{hypothesis}[Geopolitical bias]\label{h1}
Western models (Claude, GPT-4o, Grok) allocate significantly more than the Chinese model (DeepSeek) to categories related to gender equality, LGBTI+ rights and anti-discrimination, reflecting differences in training data and cultural frames.
\end{hypothesis}

\begin{hypothesis}[Housing under-allocation]\label{h2}
All four models under-allocate housing relative to the real Spanish and German social budgets, reproducing a systematic bias that is independent of the national context.
\end{hypothesis}

\begin{hypothesis}[Structural convergence]\label{h3}
The four models share a similar prioritisation structure, with health and education receiving the largest shares, suggesting a common normative frame embedded in commercial language models.
\end{hypothesis}

\begin{hypothesis}[National-context sensitivity]\label{h4}
Allocations vary between the Spanish and the German scenarios in response to the country-specific challenges described in the prompt, especially in pensions, refugee integration and energy transition, exhibiting some degree of contextual adaptation without reaching the precision of real budgets.
\end{hypothesis}

\begin{hypothesis}[Under-representation of sensitive categories]\label{h5}
No model assigns more than 2\% to politically sensitive categories such as the integration of ethnic minorities, support to migrants or unaccompanied minors, irrespective of the country, reflecting representational biases in the training data.
\end{hypothesis}

The statistical analysis is non-parametric, given that the small sample size per group ($n=6$) does not justify normality assumptions. The Kruskal--Wallis test \citep{kruskal1952use} is applied separately to each macro-area within each country, under the null hypothesis that the distributions of the four models are equal, at a significance level $\alpha = 0.05$. Whenever the test yields a significant result, pairwise Mann--Whitney U tests \citep{mann1947test} are applied with Bonferroni correction \citep{dunn1961multiple} to identify the specific pairs of models responsible for the observed divergence. The mean absolute deviation between each model's allocation vector and the OECD reference vector is computed on the twelve comparable macro-areas to quantify the global divergence between models and real budgets. Pairwise Pearson correlations on the average allocation vectors across all sub-categories complete the apparatus and characterise the similarity structure between models. The textual justifications were tokenised, lower-cased and stop-word filtered, and the resulting term-frequency tables were examined to identify the lexical regimes that each model invokes when justifying its allocations.

\section{Results}\label{sec:results}

The analysis is structured progressively. First we describe the priorities of each model and the degree to which their proposals diverge from the OECD reference structure. The differences between models are then validated statistically. We then characterise the internal consistency of each model, the similarity structure between models and their sensitivity to the national context. The section closes with a qualitative analysis of the textual justifications and a formal contrast of the five hypotheses.

\subsection*{Implicit priorities and divergence from real budgets}

\begin{figure}[ht]
  \centering
  \includegraphics[width=0.46\textwidth,keepaspectratio]{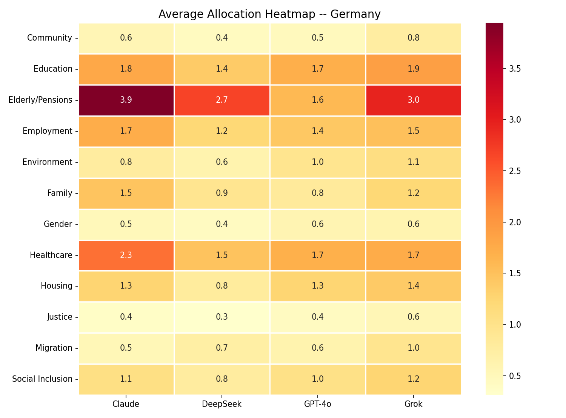}\hspace{0.02\textwidth}%
  \includegraphics[width=0.46\textwidth,keepaspectratio]{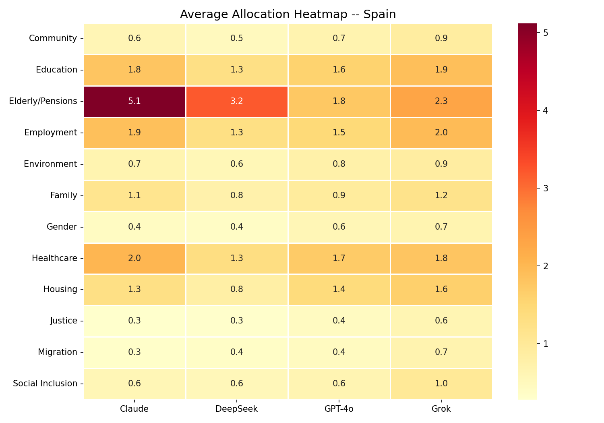}
  \caption{Heatmap of average allocations by sub-category and model. Left: Spain. Right: Germany. Each cell aggregates six independent runs per combination of model and country. The columns correspond to GPT-4o, Claude, DeepSeek and Grok.}
  \label{fig:heatmaps}
\end{figure}

The heatmaps in Figure~\ref{fig:heatmaps} represent the average allocation by sub-category and model for Spain and Germany. Four observations emerge consistently across both national contexts. The first is that Grok leads in ten of the twelve macro-areas in Spain, losing only Pensions and Health to Claude, and in eight of the twelve in Germany, with a minimum macro-area allocation of 2.5\% in Justice that exceeds the minimum of every other model. This pattern reveals an expansionist and pluralist budgetary stance, since Grok distributes resources broadly and avoids residual allocations. The second is that DeepSeek records the lowest values in most macro-areas in both countries, with the notable exception of Pensions and Older People, where the lowest allocation belongs not to DeepSeek (16.4\% in Spain; 13.7\% in Germany) but to GPT-4o (7.9\% and 8.0\% respectively). In Pensions the ordering of the models is inverted with respect to the general pattern, with GPT-4o, usually in intermediate positions, becoming the most restrictive by a wide margin. The third is that Claude concentrates the budget more pronouncedly than any other model, leading clearly in Pensions (5.1\% per sub-category in Spain) and Health (2.0\%), absorbing a disproportionate share of the budget; its lowest macro-area allocation is Justice at 0.3\%, which yields a maximum-to-minimum ratio of approximately seventeen, the largest of the four models. The fourth is that GPT-4o produces the visually most homogeneous heatmap, with sub-category values oscillating between 0.4\% in Justice and 1.7\% in Health in Spain, and with the lowest Pensions allocation (1.8\% per sub-category) combined with a standard deviation across runs of $\pm 2.35$ points, which establishes the under-allocation of Pensions in GPT-4o as systematic rather than stochastic.

\begin{figure}[ht]
  \centering
  \includegraphics[width=0.88\textwidth,keepaspectratio]{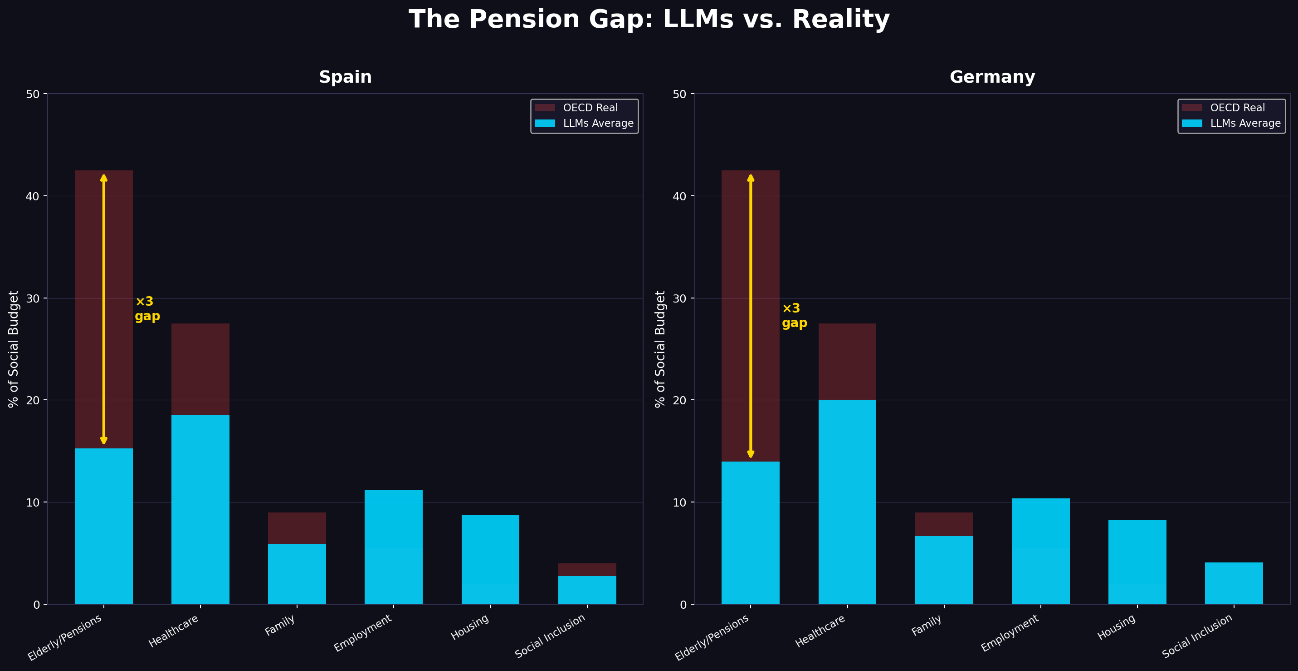}
  \caption{Average allocation by the four language models contrasted with the approximate OECD reference structure for both Spain and Germany. Pensions are under-allocated by a factor close to three, while Housing and Employment are over-allocated by factors of four and two respectively.}
  \label{fig:vs_oecd}
\end{figure}

Figure~\ref{fig:vs_oecd} contrasts the average allocation of the four models with the approximate OECD reference budget for the macro-areas where empirical comparison is available. The most striking discrepancy concerns Pensions and Older People. The OECD reference value lies between 40\% and 45\% of social spending, while the four-model average attains only 15.3\% in Spain and 14.1\% in Germany, a difference exceeding twenty percentage points and a misalignment factor close to 2.8. No individual model approaches the reference value: the closest is Claude (25.6\% in Spain; 19.7\% in Germany), which still falls more than fifteen percentage points short, while the farthest is GPT-4o (7.9\% and 8.0\%), which allocates roughly a fifth of the real share. The directionality of the error is asymmetric across macro-areas. Health is also under-allocated but to a smaller extent (average across language models $\approx 18.6\%$ versus an OECD value close to 27.5\%, a gap of $-8.9$ percentage points), as is Family ($\approx 5.9\%$ versus 9\%, a gap of $-3.1$ percentage points). On the opposite side, Housing is systematically over-allocated ($\approx 8.7\%$ versus an OECD value of approximately 2\%, a factor of 4.4) and Employment is over-allocated by a factor of two ($\approx 11.2\%$ versus 5.5\%).

The asymmetry between under-allocated and over-allocated macro-areas admits a structural interpretation. Pensions and Health are the largest fiscal inertias of European welfare systems, with budget shares that reflect decades of contributory commitments and structural healthcare expenditure. Housing and Employment, by contrast, have acquired strong salience in contemporary public discourse, particularly through the European housing crisis and the persistent visibility of youth unemployment in Mediterranean economies. The four models redistribute the budget towards the categories that dominate the public conversation rather than towards those that dominate the actual fiscal accounts. Language models behave, in this sense, as mirrors of public debate rather than of fiscal reality. The fact that the pattern is essentially identical in Spain and Germany confirms that the bias is systemic rather than country-specific.

\subsection*{Statistical validation}

\begin{figure}[ht]
  \centering
  \includegraphics[width=0.78\textwidth,keepaspectratio]{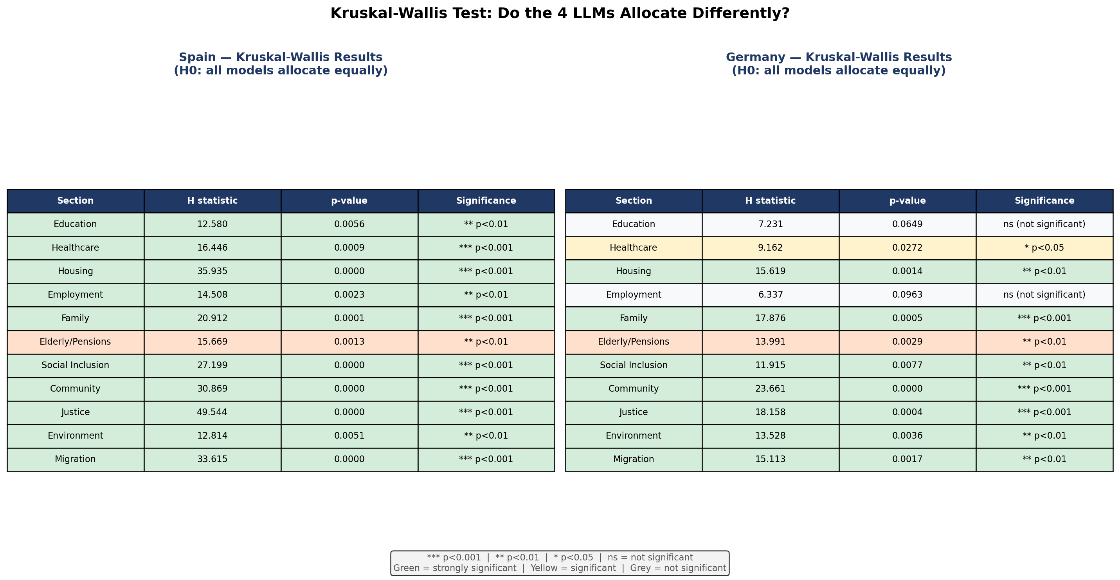}
  \caption{Kruskal--Wallis test statistics and $p$-values by macro-area and country. The null hypothesis of equality of distributions across the four models is rejected in eleven out of eleven macro-areas in Spain and in nine out of eleven in Germany.}
  \label{fig:kw}
\end{figure}

The Kruskal--Wallis test was applied to each macro-area, with the null hypothesis of equal distributional behaviour across the four models. The results are summarised in Figure~\ref{fig:kw}. In the Spanish scenario the null hypothesis is rejected in all eleven analysed macro-areas. The largest test statistics correspond to Justice ($H = 49.54$, $p < 0.001$), Housing ($H = 35.94$, $p < 0.001$), Migration ($H = 33.61$, $p < 0.001$) and Community ($H = 30.87$, $p < 0.001$), which indicates that differences between models are especially systematic in those macro-areas. It is worth noting that the macro-areas with the strongest statistical signal are those with the smallest budget weight: where the models operate in narrow ranges of 1\% to 5\%, the differences between them are proportionally more concentrated and more consistent. For Pensions the test yields $H = 15.67$ with $p = 0.001$, confirming that the gap between Claude (25.6\%) and GPT-4o (7.9\%) is statistically robust even with only six runs per model. In the German scenario, nine out of eleven macro-areas attain statistical significance, with Education ($H = 7.23$, $p = 0.065$) and Employment ($H = 6.34$, $p = 0.096$) as the only exceptions. These two macro-areas correspond to the most concrete challenges described in the German prompt, namely skilled labour shortages and the energy transition, which suggests that sufficiently specific contextual information acts as an attractor of consensus and attenuates the differences between models, though only in macro-areas with sharply defined challenges.

Post-hoc pairwise Mann--Whitney U tests with Bonferroni correction localise the specific pairs of models responsible for the observed differences. In Pensions for Spain, the most divergent pairs are Claude versus GPT-4o ($U = 555$, $p = 0.017$) and Claude versus DeepSeek ($U = 628$, $p = 0.008$). In Justice the contrasts are the sharpest of the entire experiment, with Claude exceeding GPT-4o ($U = 68$, $p < 0.001$) and GPT-4o allocating significantly less than Grok ($U = 100$, $p < 0.001$). These results confirm that the differences between models are not attributable to chance and reflect structurally distinct distributive preferences.

\subsection*{Internal consistency}

\begin{figure}[ht]
  \centering
  \includegraphics[width=0.86\textwidth,keepaspectratio]{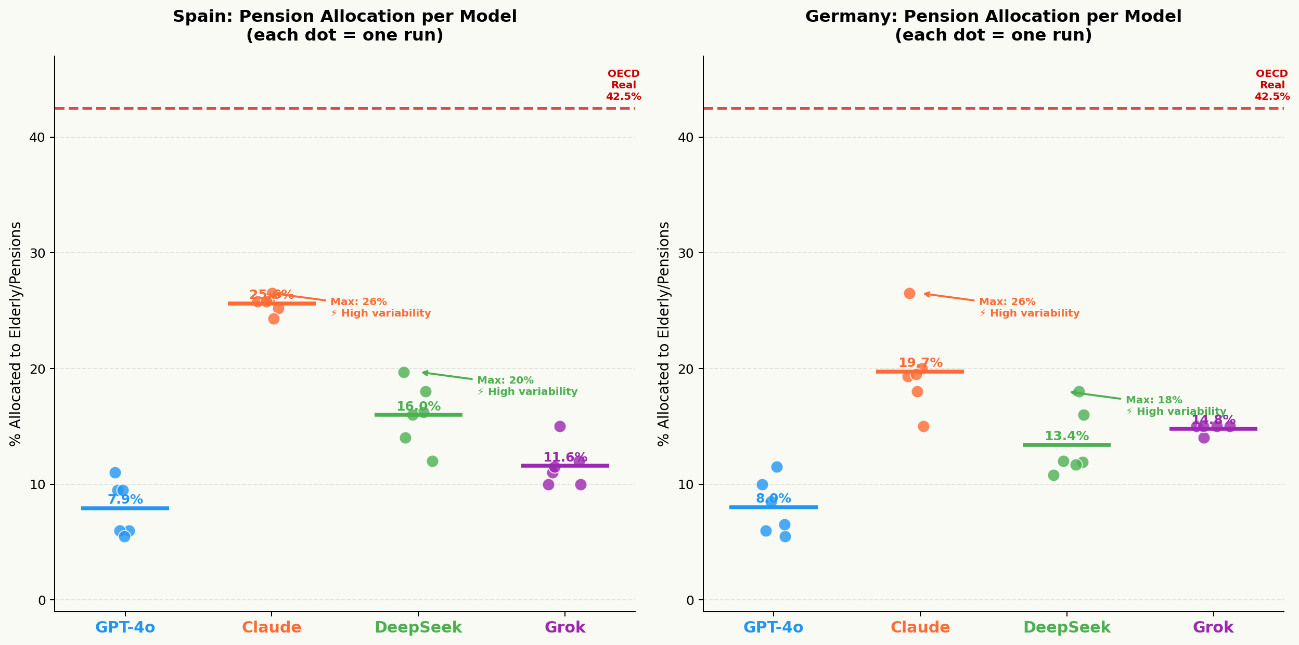}
  \caption{Distribution of the six runs per model in the macro-area of Pensions and Older People for Spain and Germany. The variance of a model is not a stable property: Claude is the most consistent in Spain and one of the most volatile in Germany, while Grok exhibits the opposite pattern.}
  \label{fig:pensions_strip}
\end{figure}

Figure~\ref{fig:pensions_strip} reports the distribution of the six runs by model in the macro-area of Pensions and Older People, where between-run variability is most pronounced. The central finding contradicts a superficial reading. Claude, despite recording the highest average allocation to Pensions in Spain (25.6\%), is also the most stable model in that macro-area, with a standard deviation of $\pm 0.76$ percentage points, the lowest of the four models. The corresponding strip plot for Claude in Spain shows the six points clustered in a compact range of 24\% to 26\%, with virtually no dispersion. Its decision to prioritise Pensions is therefore neither accidental nor volatile, but rather a systematic and robust preference repeated across all runs. DeepSeek is the most variable model in Pensions for Spain (standard deviation $\pm 2.75$), with allocations oscillating between 12\% and 20\%. In Germany the configuration changes substantially: Claude becomes the most volatile in Pensions (standard deviation $\pm 3.78$, with runs ranging from 15\% to 26\%), while Grok becomes the most stable (standard deviation $\pm 0.45$, with all runs clustered near 14.8\%). The stability of a model in a given macro-area is therefore not a fixed property of the system but depends on the interaction between its architecture and the specific context of the prompt.

The complete table of means and standard deviations reveals two further extreme cases. The lowest variance in the entire experiment for Spain corresponds to Claude in Justice (1.12\% $\pm$ 0.16), with values consistently between 0.9\% and 1.3\% across runs. The only case of nearly null variance in the whole dataset occurs for Grok in Social Inclusion for Germany (5.00\% $\pm$ 0.00), where the six runs produce identical values. At the opposite end, the largest instability of the experiment is DeepSeek in Health for Spain ($\pm 5.71$), with allocations between 7\% and 24\% depending on the run, a level of volatility that compromises the reliability of its proposals as decision support.

\subsection*{Sensitivity to national context}

\begin{figure}[ht]
  \centering
  \includegraphics[width=0.55\textwidth,keepaspectratio]{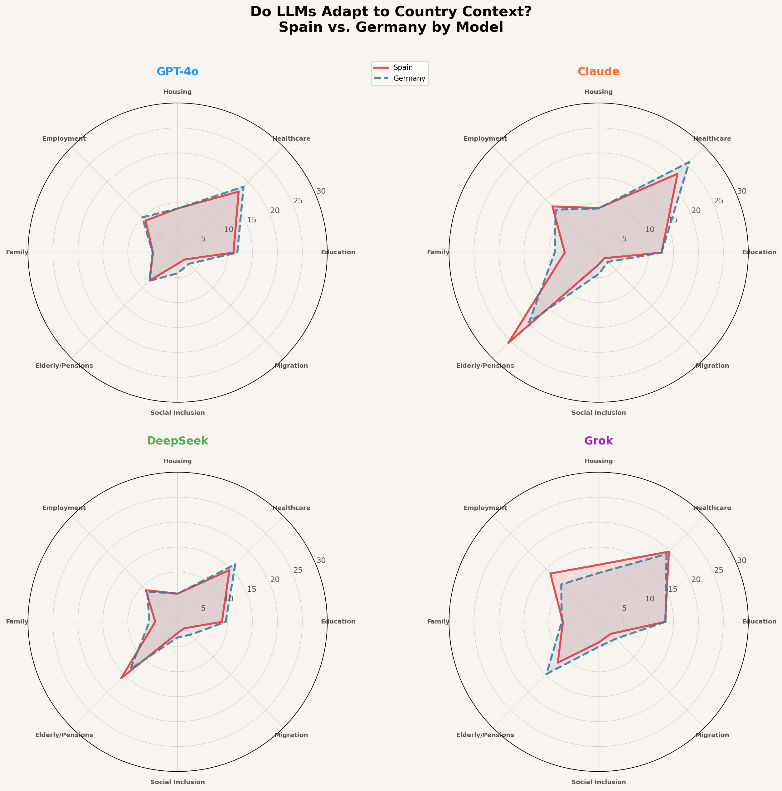}
  \caption{Adaptation of each model to the national context. The Spanish profile (continuous line) and the German profile (dashed line) overlap almost perfectly for GPT-4o and DeepSeek. Claude exhibits the largest inter-country difference, exchanging the leadership of Pensions and Health between scenarios. Grok shows a modest increase in Migration in Germany.}
  \label{fig:radar}
\end{figure}

To assess the capacity of each model to adapt to the national context, Figure~\ref{fig:radar} superimposes the Spanish profile (continuous line) and the German profile (dashed line) for each model over eight selected macro-areas. The most evident result is that GPT-4o and DeepSeek display almost identical radar profiles across countries: the two lines essentially overlap. For these models, describing the specific challenges of Spain or Germany in the prompt does not produce an appreciable change in the proposed distribution; their budgetary reasoning is, in this respect, decontextualised.

Claude is the model that adapts most strongly to the country, and does so coherently with the contextual information provided. In Spain, where the prompt highlights accelerated ageing and the underfunded long-term care system, Claude allocates more to Pensions (25.6\%) than to Health (22.4\%). In Germany, where the prompt highlights refugee integration and the energy transition alongside ageing, Claude inverts the order: Health rises to the leading position (25.7\%) while Pensions drops to second (19.7\%). This exchange of leadership between macro-areas is the largest cross-country shift observed in the experiment. Grok exhibits a perceptible difference in Migration (3.4\% in Spain versus 4.8\% in Germany), consistent with the description of more than two million refugees in the German prompt, although the absolute magnitude of the shift remains modest. In no case do the models reproduce the structural quantitative differences between the two welfare systems.

\subsection*{Pairwise similarity between models}

The correlation matrices of average allocation vectors across all sub-categories reveal the most counterintuitive result of the experiment. The two models with the highest structural similarity are not two Western systems, but Claude (Anthropic, United States) and DeepSeek (DeepSeek AI, China), with $r = 0.99$ in Spain and $r = 0.97$ in Germany. This contradicts the geopolitical hypothesis articulated in Hypothesis~\ref{h1}, which predicted a Western versus Chinese principal axis. The true axis of differentiation is the contrast between concentration, exemplified by Claude and DeepSeek, which place disproportionate weight on a few macro-areas, and dispersion, exemplified by GPT-4o, which distributes more uniformly across the taxonomy. GPT-4o is the least correlated model with the others (for example, $r = 0.80$ with Claude and $r = 0.76$ with DeepSeek in Spain). The geographical origin of a model does not predict its distributive preferences; rather, what is predicted is whether the model concentrates or disperses its allocation.

\subsection*{Qualitative analysis of justifications}

The lexical analysis of the textual justifications reveals two distinct rhetorical regimes. GPT-4o organises its reasoning around absolute monetary magnitudes: the term ``billion'' appears with very high frequency (approximately four hundred occurrences in each country), and justifications are typically built around statements of the form ``X billion ensures'' rather than around the policy goals of the intervention. Grok shares this orientation, although less radically. Claude and DeepSeek exhibit a different vocabulary, organised around terms such as ``support'', ``programs'', ``social'', ``care'' and ``integration''. These two models reason in terms of objectives, that is, what an allocation is intended to achieve, rather than in terms of quantitative magnitudes. This programmatic orientation aligns with the high mutual correlation between Claude and DeepSeek ($r = 0.99$), since they arrive at similar distributions on the basis of arguments that prioritise expected social impact. In the German scenario, Grok is the only model that introduces ``refugee'' and ``ageing'' among its most frequent terms, which is coherent with its larger inter-country differentiation. DeepSeek maintains ``social'' as the leading term in both scenarios, reflecting a frame oriented towards collective cohesion that does not vary with the country.

\subsection*{Formal contrast of hypotheses}

We now examine the five hypotheses formulated in Section~\ref{sec:methods}. Hypothesis~\ref{h1}, on geopolitical bias, is refuted in its structural dimension and only partially confirmed in its categorical dimension. A difference in Gender Equality in the predicted direction is observed, with DeepSeek allocating less than the Western models, but the correlation matrix invalidates the central premise: Claude, the Western model that should differ most from DeepSeek according to the hypothesis, is in fact the model most similar to it ($r = 0.99$ in Spain). The principal axis of differentiation is not geopolitical. Hypothesis~\ref{h2}, on housing under-allocation, is refuted: the four models systematically over-allocate housing (average across language models of $\approx 8.7\%$ in Spain and $\approx 8.2\%$ in Germany versus an OECD value of approximately 2\%, a factor of 4.4). The directionality of the bias is opposite to the expected one and is coherent across all four models and both countries. Hypothesis~\ref{h3}, on structural convergence, is partially confirmed: Health and Education indeed appear among the top priorities of all models, but the gap in Pensions (GPT-4o 7.9\% versus Claude 25.6\%) precludes full convergence, and Kruskal--Wallis rejects equality in ten out of eleven macro-areas in Spain and nine out of eleven in Germany. Hypothesis~\ref{h4}, on national-context sensitivity, is only weakly confirmed: only Claude shows appreciable adaptation, with the exchange of leadership between Pensions and Health, while GPT-4o and DeepSeek are essentially insensitive to the change of context. Grok shows a modest adaptation in Migration. None of the models reproduces the structural quantitative differences between the two welfare systems. Hypothesis~\ref{h5}, on the under-representation of politically sensitive categories, is confirmed: Migration and Integration receives an average of 2.1\% in Spain and 3.5\% in Germany; at the sub-category level, allocations such as care of unaccompanied minors, anti-trafficking programs and the integration of ethnic minorities never exceed 1\% systematically across any of the forty-eight runs. The under-representation is robust and consistent with the literature on representational bias in training data \citep{bender2021stochastic}.

In summary, the data confirm the general hypothesis of the work: the four models exhibit an implicit social policy that diverges systematically from real European budgetary structure and does not depend on the national context.

\section{Discussion}\label{sec:discussion}

Three implications follow from the empirical findings. The first concerns the structural bias of language models in social policy. The four models, despite their architectural differences and the geographical diversity of their developers, share a systematic distortion of the European welfare budget structure. Pensions are under-allocated by a factor close to three, Health is under-allocated by approximately one third, while Housing is over-allocated by a factor of four and Employment by a factor of two. The systematic nature of this pattern, replicated across countries and models, indicates that the bias originates in the training data rather than in any particular alignment strategy. Language models behave as mirrors of contemporary public discourse, in which the housing crisis and youth unemployment have far greater discursive presence than contributory pensions and structural healthcare spending. This finding qualifies the existing literature on the political preferences of language models: while \citet{rozado2024political} documents a left-leaning inclination after alignment, our results show that the distributive bias is not reducible to a left-right axis. A left-leaning model could be expected to over-allocate to redistributive categories such as pensions and unemployment benefits; we find the opposite for pensions and a more nuanced pattern overall, which suggests that the salience of categories in training corpora is at least as important as the ideological orientation of the alignment process.

The second implication concerns the limitations of geopolitical framings of language models. Our results indicate that the country of origin of a model is not a reliable predictor of its distributive preferences. The most similar pair in the experiment, Claude and DeepSeek, combines a United States-based model and a Chinese model with $r = 0.99$ in Spain. The principal axis of variation is not Western versus Chinese, but concentrated versus dispersed allocation, which is a property of the alignment strategy and of the structure of the training data rather than of the geopolitical context. This nuances the public debate over cultural and ideological dimensions of artificial intelligence and suggests that comparative evaluations should focus on architectural and methodological dimensions rather than on national labels.

The third implication concerns sensitivity to context. The weak adaptation of GPT-4o and DeepSeek across the Spanish and German scenarios, despite explicit contextual cues in the prompt, raises substantive doubts about the suitability of these systems as decision support in policy analysis where country-specific structural features are critical. Only Claude exhibits appreciable adaptation, and even so it does not reproduce the structural differences between the two welfare systems. This is consistent with the observation by \citet{santurkar2023whose} that language models tend to reflect averaged opinions of the populations represented in their training data, which dilutes country-specific signals. The high internal consistency of these models, in particular in Claude's allocation to Pensions in Spain or Grok's allocation to Social Inclusion in Germany, indicates that their distributive choices are stable preferences rather than stochastic outputs, but this stability does not translate into faithful contextual sensitivity.

The implications for the use of language models in public deliberation are correspondingly delimited. Used as decision support in social budgeting, current commercial systems would systematically reproduce the structural biases identified here, and the influence on human decisions documented by \citet{fisher2024biased} suggests that such bias would propagate into the deliberative process even when users are unaware of it. The appropriate role of these systems is therefore as a complement to expert deliberation, used to generate alternative scenarios or to surface arguments not initially considered, but never as an autonomous allocator. The design of correction mechanisms, whether through fine-tuning on real budgetary references, structured prompting that injects empirical benchmarks, or post hoc calibration on the OECD distribution, constitutes a natural research line that would translate the diagnostic results of this work into operational tools.

\section{Conclusions and future work}\label{sec:conclusion}

This work has examined how four commercial large language models distribute a real social budget across twelve macro-areas of public expenditure for Spain and Germany, with the aim of characterising their implicit social policy and assessing their potential as decision support in public budgeting. The principal finding is that all four models exhibit a systematic departure from the real European budget structure, with a severe under-allocation to Pensions, a moderate under-allocation to Health and Family, and a marked over-allocation to Housing and Employment. The bias is replicated across both countries and reflects, plausibly, the salience structure of contemporary public discourse rather than the fiscal reality of European welfare states. The second finding is that the differences between models are statistically robust, with Kruskal--Wallis rejecting equality in almost all macro-areas, yet the principal axis of differentiation is not geopolitical: Claude (United States) and DeepSeek (China) are the most similar pair, while GPT-4o is the least correlated with the rest. The relevant axis is concentration versus dispersion of the budget. The third finding is that sensitivity to the national context is limited and asymmetric: only Claude adapts appreciably, while GPT-4o and DeepSeek produce essentially identical distributions in both scenarios.

Several directions for future work follow naturally. A first extension would broaden the geographical scope by including countries with very different welfare structures, such as the Nordic economies with social spending close to 30\% of gross domestic product, and emerging economies with much lower social spending, in order to determine whether the biases observed here are universal or specific to a European context. Open-weight models such as Llama and Mistral should be included to test whether the biases are characteristic of commercial systems with strong alignment by reinforcement learning from human feedback or extend to systems with less intervention in the training process. A second extension would study the systematic effect of prompt design, by injecting OECD reference data, by adding explicit budgetary constraints, or by varying the language of the prompt; the persistence of biases under more informative prompts would reinforce the diagnosis that the bias is structural. A third extension would examine the temporal evolution of these biases across successive releases of each model, in order to assess whether developers are actively correcting them. A fourth extension would refine the analysis of the textual justifications by applying topic modelling and automated ideological classification to the full text of the justifications, in order to characterise the normative frames invoked by each model with greater precision. A fifth extension, of particular societal relevance, would experimentally measure the impact of exposure to language-model-generated allocations on the budgetary preferences of human participants, connecting our findings to the influence literature initiated by \citet{fisher2024biased}. Finally, the design and evaluation of explicit correction mechanisms, whether through fine-tuning on real budgetary references or through structured prompting that incorporates empirical benchmarks, would translate the diagnostic results of this work into operational tools for responsible deployment of language models in public decision support.

\bibliographystyle{plainnat}
\bibliography{references}

\end{document}